# Interaction of Polymer of Intrinsic Microporosity PIM-1 with explosive analytes at the molecular level: Combined experiment and computational modelling

Salam Mohammed[1,2], Edward B. Ogugu[2], Ramakant Sharma[2], Dominic Taylor[3], Graeme Cooke[4], Neil McKeown[3], Glib Baryshnikov[5], Hans Ågren[6,7], Ifor D.W. Samuel[2], Graham A. Turnbull[2]

1. Swedish EOD and Demining Centre-SWEDEC, Swedish Armed Forces, SE-575 28 Eksjö, Sweden.
2. Organic Semiconductor Centre, School of Physics and Astronomy, University of St Andrews, St Andrews, KY16 9SS, UK.
3. School of Chemistry, University of Edinburgh, Edinburgh EH9 3FJ, UK.
4. School of Chemistry, University of Glasgow, Glasgow G12 8QQ, UK.
5. Department of Science and Technology (ITN), Linköping University, SE-581 8 Linköping, Sweden.
6. Department of Physics and Astronomy, X-ray Photon Science, Uppsala University Box 516, 751 20 Uppsala, Sweden.
7. Faculty of Chemistry, Wroclaw University of Science and Technology, Wyspiańskiego 27, PL-50370 Wroclaw, Poland

*Author e-mail address: gat@st-andrews.ac.uk and salam.mohammed@mil.se*

## Abstract

This work investigates the molecular-level interactions of a fluorescent microporous polymer (PIM-1) with nitroaromatic explosives, in the context of thin film explosive sensors. Thin films of the PIM-1 were exposed to 2,4-dinitrotoluene (DNT) and 2,4,6-trinitrotoluene (TNT), and their steady-state absorption and emission spectra measured. For comparison, the response of PIM-1 to non-explosive molecules such as benzene (BN) was also explored. Complementary electronic-structure calculations were used to predict absorption and emission spectra and to determine binding energies for the PIM-1–analyte complexes. The calculations agree well with experiment and reveal that association of nitroaromatic analyte molecules with PIM-1 alters the energy levels and the arrangements of frontier orbitals, indicating significant molecular interactions. Calculations show that the electronic properties and photo-excited electron transfer can be described by interaction with a single repeat unit of the polymer. The molecular binding, however, involves interaction with at least three repeat units, with the DNT/TNT molecule binding into a pocket in the contorted structure of the microporous polymer. Together, the experimental and theoretical results demonstrate that PIM-1 is a promising platform for selective nitroaromatic detection and provide molecular design principles that could improve sensitivity and selectivity in future sensor materials.

The detection of trace amounts of hazardous chemicals presents a significant and critical challenge across a variety of fields, including homeland security, environmental monitoring, and humanitarian landmine clearance efforts[1]. These applications require highly sensitive and reliable methods for identifying harmful substances at very low concentrations, particularly in complex environments where conventional detection techniques may be less effective[2,3]. Among the most promising approaches for achieving this level of sensitivity are fluorescent polymer sensors, which capitalize on an optical quenching mechanism activated in the presence of vaporized explosive molecules such as TNT[4–6]. These sensors provide an attractive solution for detecting concealed explosives, chemical residues, or forensic traces, offering a remarkable level of sensitivity and specificity that is essential in applications where every detection is vital to safety and security[7,8].

The fundamental principle behind the operation of these fluorescent polymer sensors is the fluorescence quenching process[9]. This process occurs when a photoexcited polymer chain transfers an electron to an adjacent nitroaromatic molecule, such as those present in explosive vapors[10]. This electron transfer leads to a reduction in the light emission from the sensor film, which is the basis for detection. As a result, the system operates on a molecular scale, enabling the detection of even extremely low concentrations of hazardous chemicals[11]. When explosive molecules are present in the vapor surrounding the fluorescent polymer film, they are first absorbed into the polymer matrix. Once absorbed, these molecules can influence the sensor's optical properties in several ways[12]. For example, the aforementioned change in the fluorescence intensity, but also a change in fluorescence wavelength, or they may alter the optical absorption spectrum, which can be measured using colorimetric sensing techniques[13].

One of the key features that enhances the sensitivity and speed of detection in these systems is the strong non-covalent interactions between the fluorophore film and the absorbed analyte molecules[11]. These interactions can lead to the accumulation of even weak vapor concentrations from the headspace above the sensor, which effectively increases the sensor's response[14]. As a result, the detection of hazardous chemicals becomes faster and more sensitive, even in environments with low vapor concentrations. Additionally, these interactions can help provide a distinctive signature for specific analytes, enabling selective detection of particular explosives or chemicals when the explosive molecule is desorbed from the fluorescent polymer film[15]. This ability to selectively identify specific analytes further enhances the value of fluorescent polymer sensors in sensitive and high-stakes applications such as bomb detection or environmental monitoring.

Polymers of intrinsic microporosity (PIMs) were originally coined and developed by one of the present authors[16]. They represent a promising class of materials that have garnered significant attention due to their unique structural properties and potential applications in sensing technology[17,18]. PIMs are macromolecules with highly contorted, rigid molecular structures, which create an amorphous, porous network on the nanometer scale[19]. This structure allows them to possess a high surface area and high gas permeability, making them ideal candidates for use in gas separation and filtration systems[20,21]. More importantly, the ability to control molecular interactions through their molecular design is a key feature that makes PIMs particularly attractive for advanced sensing applications[6].

Certain PIMs are fluorescent, which provides an opportunity to significantly enhance fluorescence-based sensing techniques. By carefully tailoring the design of these materials, it is possible to create sensors with optimized sensitivity, speed, and selectivity for a range of chemical threats. For example, their high surface area and tunable porous structure allow for rapid analyte absorption, which accelerates the overall detection process. Furthermore, by adjusting the molecular design, it is possible that PIMs may offer routes to enhance the polymer's interaction with specific analytes, enabling the development of selective sensors that can differentiate between different chemical compounds[22]. It is therefore very important to better understand the interaction between the PIM matrix and analyte molecules of interest at the molecular scale. To this end, we combine spectroscopic measurements and *ab initio* calculations of the interaction of the prototypical polymer PIM-1 and common nitroaromatic molecules.

In the present work, we show that interaction between PIM-1 and explosive analytes (DNT and TNT) causes a spectral shift in absorption and strong quenching of the fluorescence due to the photoexcited electron transfer in the bound donor-acceptor complex. We visualize this process within the density functional theory (DFT) simulations to confirm a photo-excited electron transfer to the analyte. Furthermore, we estimate the binding energy experimentally from thermal release of bound analyte molecules at elevated temperature. We find that the non-covalent binding of DNT and TNT to PIM-1 can be predicted by DFT simulations, and involves an electrostatic interaction with multiple repeat units of the polymer.

**Materials and Methods**

PIM-1 was synthesised by the double nucleophilic aromatic substitution reaction between 5,5',6,6'-tetrahydroxy-3,3,3',3'-tetramethyl-1,1'-spirobisindane and tetrafluoroterephthalonitrile in the presence of potassium carbonate as a base[23,24]. This yielded a bright yellow powder that was purified by reprecipitation from chloroform solution into methanol: the purity of the sample was confirmed by NMR spectroscopic characterisation and gel permeation chromatography (see SI 1).

Microporous films of PIM-1 were fabricated by spin-coating a 20 mg/mL PIM-1 solution in chloroform at 2000 rpm for 60 seconds on 12 mm diameter fused silica substrates (UQG optics), resulting in a film thickness of approximately 180 nm, as measured by spectroscopic ellipsometry. Prior to spin coating, the substrates were cleaned ultrasonically for 10 minutes in acetone, followed by isopropanol, then dried in a nitrogen stream, and plasma-ashed in 100% oxygen (Plasma Technology MiniFlecto) for 3 minutes. The PIM-1 films were subsequently doped with explosive analytes DNT and TNT by drop-casting 20 μL of 1 mM solutions in acetonitrile onto the polymer films, yielding masses of 3.64 μg and 4.45 μg for DNT and TNT, respectively. The solutions were allowed to evaporate, leaving behind molecules of analytes adsorbed in the films before steady-state measurements. The response of PIM-1 to non-explosive molecules was also explored by drop casting 20 μL of benzene (BN).

UV-VIS absorption (using Jasco 770 UV-Vis-NIR Spectrophotometer) and fluorescence spectra (using Edinburgh Instruments FS5) of the films were measured before and after analyte exposure, and also after thermal desorption of the sorbed analyte. The thermal desorption data were used to estimate the binding energy of DNT, as discussed in the results and discussion section. For this experiment, the PIM-1 film was loaded with 16.39 µg of DNT and heated using a hot plate for 3 minutes at each of a series of temperatures starting from 40 ℃, and increasing in steps of 10°C. We assume that the loaded 16.39 µg of DNT is sorbed into the PIM-1 film, as polymeric materials are known to swell, allowing analyte molecules to be sorbed into the thin films [25,26]. And the thermal desorption is largely from the bulk of the PIM-1 film. The sample was heated in a fume hood to enable the safe extraction of DNT vapors from the lab. After each heating step, an absorption measurement was made to determine the amount of analyte lost. The film was then removed from the spectrophotometer and heated at a higher temperature before the next absorption measurement, and the process was repeated up to 200 ℃. The characteristic optical responses arise from differences in polymer-analyte binding and photoinduced electron transfer quenching. To relate these processes to sensor response we made a combined experimental and theoretical study of explosives sensing using the microporous polymer PIM-1.

A molecular model of PIM-1 (Fig. 1a) was used for calculations of optical properties of absorption and emission, represented for electronic level calculations by a single repeat unit 1U-PIM-1 of the polymer (Fig. 1b). Thus, the spiro-centre and methyl groups of PIM-1 have been omitted and the terminal substituted by a hydrogen atom[27]. However, to accurately calculate the binding energy ($\Delta E$), we included the contorted PIM-1 structure with multiple repeat units up to ten oligomer units (10U-PIM-1) of PIM-1. Full geometry optimization for both isolated 1U-PIM-1 and PIM-1 complexes (1U-PIM-1+DNT, 1U-PIM-1+TNT and 1U-PIM-1+BN) were then performed at the density functional theory (DFT) level, using the hybrid B3LYP exchange–correlation functional[28] in conjunction with the 6-31g(d) basis set for all cases (Fig. 1a-e and SI 2). All calculations were performed using the Gaussian 16 program[29]. The absorption spectra and fluorescence emission characteristics of 1U-PIM-1 and PIM-1 complex with analyte molecules (DNT, TNT and BN) were calculated by using time-dependent (TD) DFT[30] at B3LYP/ 6-31g(d) level. However, for the calculations of $\Delta E$, we applied B3LYP-gCP-D3/6-31G(d), since the B3LYP functional has known limitations in accurately modelling non-covalent interactions like dispersion forces[31–35] (see the comparison between the two approaches in SI 10).

The optimized $\Delta E$ between a DNT molecule and four different oligomeric configurations of PIM-1 (1U-PIM-1, 2U-PIM-1, 3U-PIM-1 and 10U-PIM-1) were also investigated using B3LYP-gCP-D3/6-31G(d). The $\Delta E$ between XU-PIM-1 and DNT is defined as:

$$\Delta E = E(\text{XU-PIM-1+DNT}) - [E(\text{XU-PIM-1}) + E(\text{DNT})] \qquad (1)$$

Where $E$ (XU-PIM-1+DNT), $E$ (1U-PIM-1) and E (DNT) represent the energy of, the XU-PIM-1+DNT, XU-PIM-1 and the DNT molecule, respectively.

In order to estimate the corresponding experimental desorption energy, $E_d$ during the step-wise thermal desorption of DNT from the PIM-1 film, we apply a modified Arrhenius equation to the experimental data. Herein we derive the modified Arrhenius equation starting from the Arrhenius equation[36]:

$$k = k_0 e^{-\frac{E_a}{RT}} \quad (2)$$

Where $k$ is the specific reaction rate, $k_0$ is the frequency or pre-exponential factor, $T$ is the absolute temperature, R is the ideal gas, and $E_a$ is the activation energy. The absorbance, $A$, is proportional to the concentration of quenchers $[Q(t)]$ which follows the time dependence:

$$\frac{d[Q(t)]}{dt} = -k[Q(t)]. \quad (3)$$

The solution to (3) is

$$[Q(t)] = [Q(t=0)]\exp^{(-kt)}. \quad (4)$$

We assume that the film is heated at a specific temperature during time interval ($t_0 \to t_0 + \Delta t$). For each such heating step:

$$[Q(t_0 + \Delta t)] = [Q(t_0)]\exp^{(-k\Delta t)}, \quad (5)$$

$$\frac{[Q(t_0)]}{[Q(t_0 + \Delta t)]} = \frac{A(t_0)}{A(t_0 + \Delta t)} = \exp^{(k\Delta t)}. \quad (6)$$

Thus, the specific reaction rate is given by

$$\frac{1}{\Delta t} \ln\left(\frac{A(t_0)}{A(t_0 + \Delta t)}\right) = k. \quad (7)$$

Substituting $k$ from (2) into (7) gives

$$\frac{1}{\Delta t} \ln\left(\frac{A(t_0)}{A(t_0 + \Delta t)}\right) = k_0 e^{-\frac{E_d}{RT}} \quad (8)$$

$$\ln\left\{\ln\left(\frac{A(t_0)}{A(t_0 + \Delta t)}\right)\right\} = -\frac{E_d}{RT} + \ln(\Delta t k_0) \quad (9)$$

For a series of step-wise heating steps of increasing temperature, each heating step of the series can be written as:

$$\ln\left\{\ln\left(\frac{A(t_i)}{A(t_i + \Delta t)}\right)\right\} = -\frac{E_d}{RT} + \ln(\Delta t k_0) \qquad (10)$$

In the present work we use eqn. (10) to estimate the desorption energy, $E_d$.

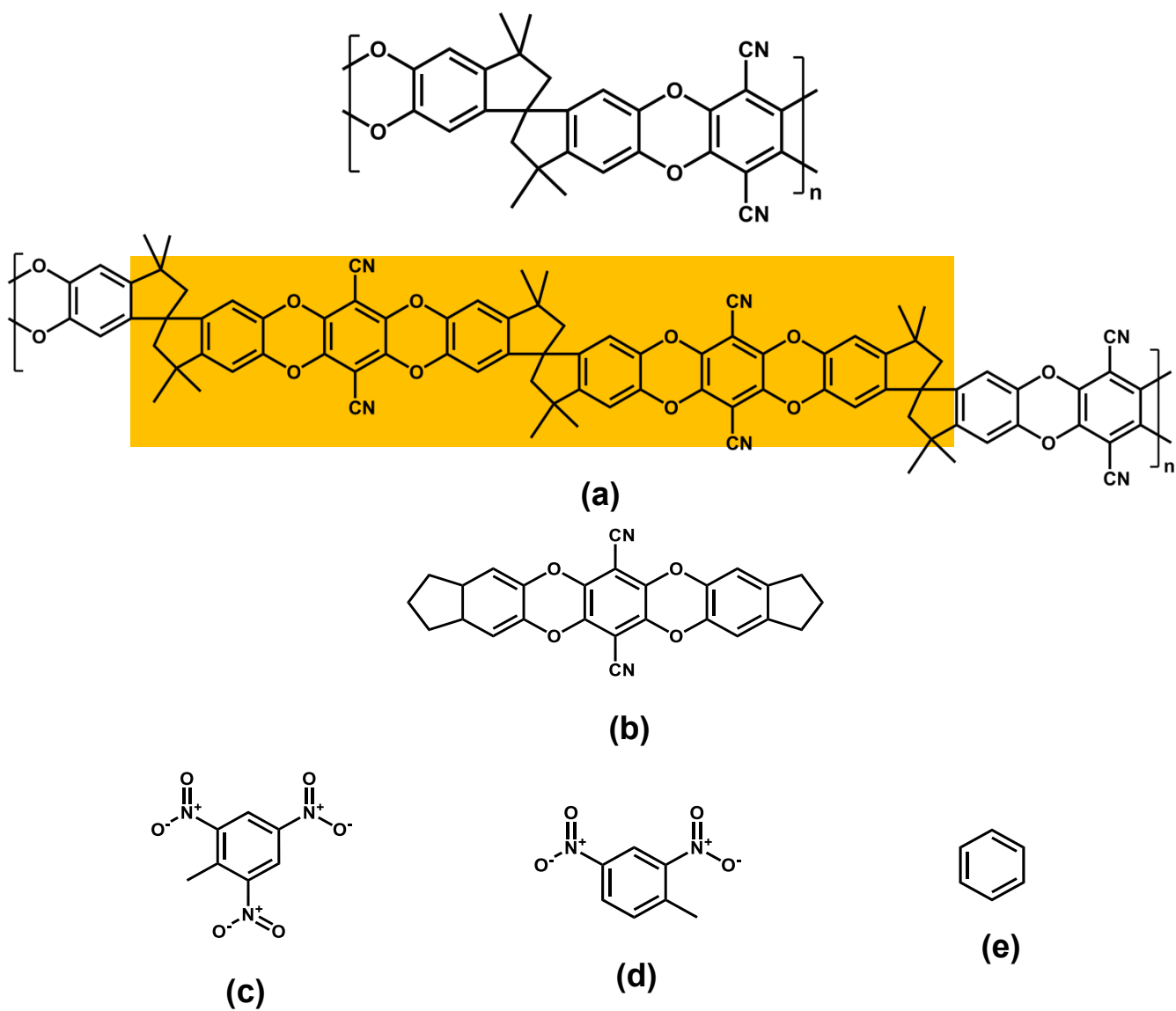

**Figure 1** – Molecular structure of the (a) PIM-1 (top) and 3 repeat units of PIM-1 (bottom), (b) molecular model of fragment of one unit of PIM-1 (1U-PIM-1), (c) DNT, (d) TNT and (e) BN.

**Results and Discussion**

The experimental and theoretical absorption and emission spectra for PIM-1 and PIM-1 complex (PIM-1+DNT, PIM-1+TNT, and PIM-1+BN) are presented in the upper and lower panels of Fig. 2, respectively. The experimental UV-Vis spectra for these four cases (PIM-1, PIM-1+DNT, PIM-1+TNT and PIM-1+BN) are dominated by three broad structureless absorption bands with a maximum around 240 nm and one comparatively weak shoulder to the main peak around 300 nm and a slightly broader band around 430 nm, respectively. In the presentation of the theoretical spectra (of 1U-PIM-1, 1U-PIM-1+DNT, 1U-PIM-1+TNT and 1U-PIM-1+BN), we have included bars to represent the oscillator strengths as well as provided a line profile obtained by an application of a Gaussian line broadening. Both the experimental measurements and the theoretical calculations show that the bands around 240 nm and 300 nm for all cases overlap without a significant change in the spectra from addition of the analyte. We note however that the experimental spectra for the band around 430 nm is red-shifted by analyte addition of DNT (PIM-1+DNT) and TNT (PIM-1+TNT), causing a shift about 10 nm compared to the isolated PIM-1 and PIM-1+BN (upper inset). These spectral changes were also predicted by the DFT calculations (lower inset) which shows a compelling overall agreement between the theoretical and experimental spectra.

The experimental and theoretical calculations of the emission spectra for PIM-1 and PIM-1 complexes (PIM-1+DNT, PIM-1+TNT and PIM-1+BN) are also presented in Fig. 2 in the wavelength region between 450 to 700 nm. Exposure to DNT or TNT leads to a strong quenching of the fluorescence, but without significant change in spectrum (SI 4). It is well known that the observed luminescence is a result of recombination of the singlet exciton state. Specifically, excited, delocalized π electrons (π*) transition to the ground state (π), releasing energy as light. However, interactions between the CN group of PIM-1 with nitroaromatic analyte (see SI 5) can give rise to photoinduced electron transfer to the nitroaromatic molecule, competing with the radiative transition, and hence leading to a reduction in luminescence. The TD-DFT calculations provide additional visualisation for this process in electronic spectra of the PIM-1 complex. The presence of the nitrate in *e.g.* DNT makes it electron-deficient so that when the DNT is in contact with the electron-rich PIM-1, photoinduced electron transfer occurs between the lowest unoccupied molecular orbital (LUMO) of the polymer and the LUMO of the analyte molecule with lower energy state. As shown in Fig. 3a the LUMO orbital level of the 1U-PIM-1 is -2.20 eV which is higher than the corresponding LUMO of the DNT (-3.00 eV) or TNT (-3.50 eV). This may suggest that the excited electrons on the LUMO orbital of 1U-PIM-1 could transfer to the LUMO of *e.g.* DNT and as a result, the fluorescence of PIM-1 complex is quenched. This is however not observed in the case of the BN analyte molecule. The LUMO orbital energy of BN is -0.10 eV which is higher than the LUMO of PIM-1 (-2.20 eV). This prohibits the formation of a charge-transfer state between PIM-1 and BN. Furthermore, calculation of the HOMO and LUMO orbitals of the complexes 1U-PIM-1+DNT and 1U-PIM-1+TNT show that the LUMO is located on the DNT and TNT respectively, while the HOMO is on 1U-PIM-1. In contrast, the LUMO orbital of complex 1U-PIM-1+BN remains unchanged compared to the LUMO of an isolated 1U-PIM-1 (Fig. 3b and SI 6).

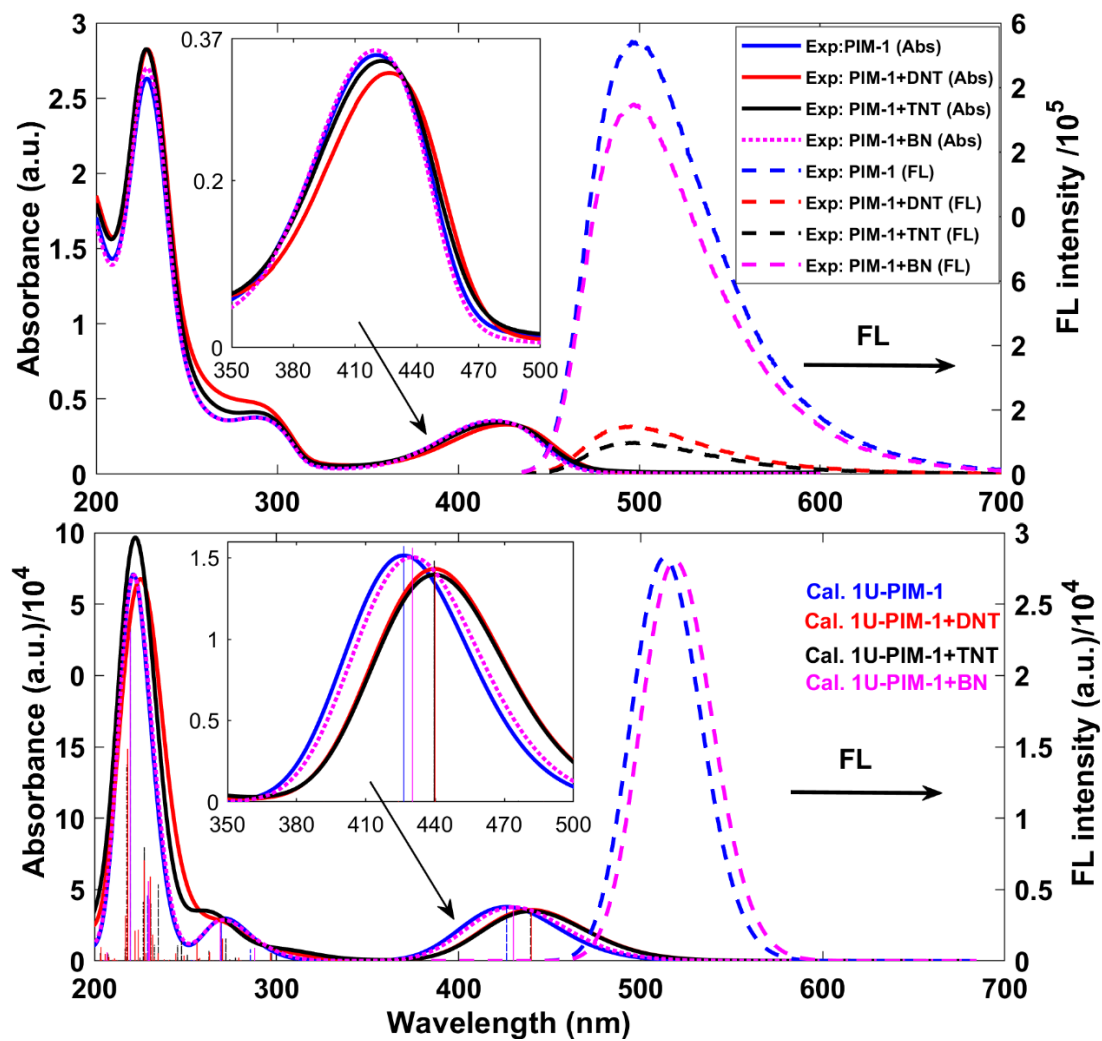

**Figure 2** – Experimental (upper panel) and calculated (lower panel) shows the absorption spectra (solid/dotted lines) and emission spectra (dashed lines) of the investigated molecules, respectively. The calculated spectra of 1U-PIM-1, 1U-PIM-DNT, 1U-PIM-1+TNT and 1U-PIM-1+BN (SI 2) are based on the electronic oscillator strength distribution, broadened by Gaussian line profiles (lower panel). The insets show expanded measured and calculated spectra in the wavelength between 350 nm to 500 nm, respectively.

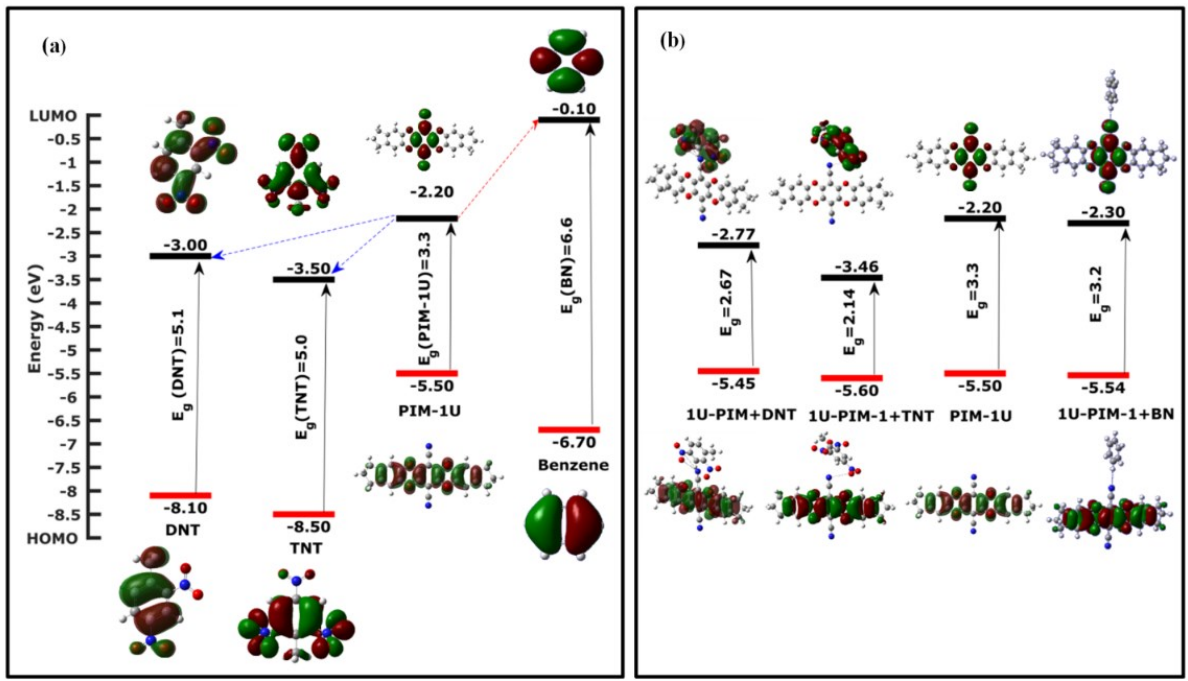

**Figure 3** – (a) Frontier molecular orbitals of HOMO and LUMO of DNT, TNT, PIM-1 and BN, respectively (b) shows HOMO and LUMO of 1U-PIM-1+DNT, 1U-PIM-1+TNT and 1U-PIM-1+BN, respectively.

We next consider the changes relating to a strong non-covalent binding interaction between PIM-1 and *e.g.* DNT.

Fig. 4a shows the change in the absorbance of the film due to 16.39 µg of DNT dispersed in a PIM-1 film of thickness 180 nm (obtained after subtracting the absorbance due to PIM-1), and the subsequent thermal desorption of the DNT following the heating procedure described in the materials and methods section. By measuring the change in absorbance at 241 nm during thermal desorption of DNT, we can estimate the desorption energy $E_d$ by applying the modified Arrhenius equation (10) to the experimental data in Fig. 4a. The observed reduction in the absorbance peaks at 241 nm can be attributed to DNT molecules being released from the film following each heating step. The DNT absorbance remains approximately constant until the film temperature is increased to 70 ºC. The drop in absorbance is initially relatively small, even when heated at 100°C, indicating the strong binding interaction between DNT and PIM-1, which can only be overcome at higher temperatures.

The negative and positive peaks at 400 nm and 450 nm correspond to the spectral shift of this PIM-1 absorption band in the presence of DNT. Interestingly, these reduce and are absent when the film is heated to 150 ºC and 200 ºC, indicating that the PIM-1 film returns to its pristine state. Fig. 4(b) shows a graph of the modified Arrhenius plot (10), used to estimate $E_d$ of the DNT thermal desorption data of Fig. 4 (a). A linear fit to the absorbance data obtained from 70

°C to 140 °C (corresponding to $2.9 \times 10^{-3}$ K$^{-1}$ to $2.4 \times 10^{-3}$ K$^{-1}$ on the $1/T$-axis, respectively) gives a desorption energy of 73 kJ/mol.

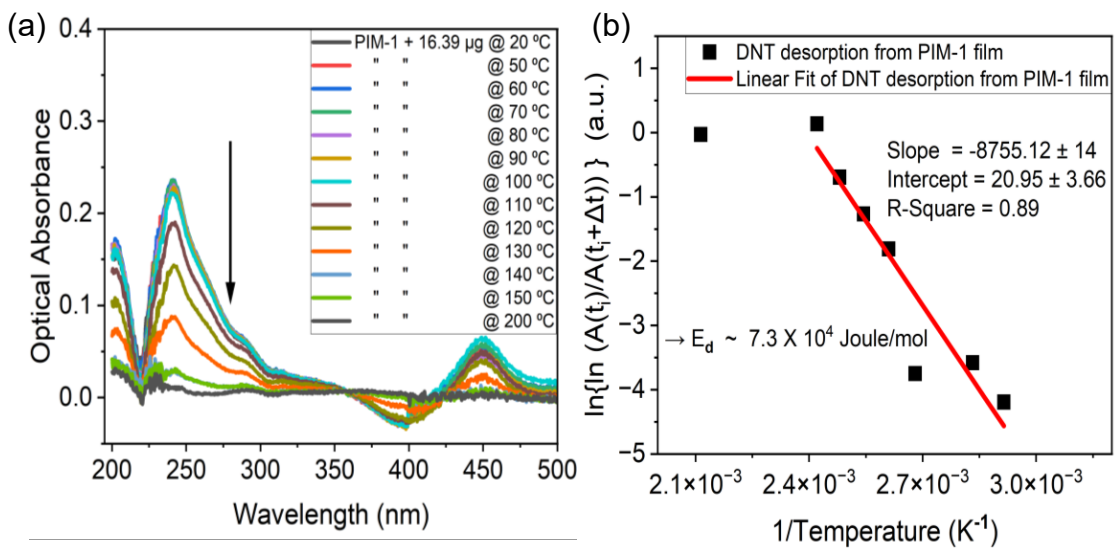

**Figure 4** – Thermal desorption of DNT from PIM-1 film: (a) Optical absorbance of 16.39 µg DNT in PIM-1 matrix, obtained after subtracting the absorbance due to pristine PIM-1, and subsequent drop in the absorbance with increase in temperature from 50 °C to 200 °C. (b) Natural log of change in absorbance against the inverse of temperature for DNT in PIM-1 film.

We also calculated the binding energy ($\Delta E$), applying the B3LYP-gCP-D3/6-31G(d) functional, for PIM-1 in interaction with DNT, using four different configurations of molecular model (1U-PIM-1+DNT, 2U-PIM-1+DNT, 3U-PIM-1+DNT and 10U-PIM-1+DNT as shown in SI 7.

The representations of calculated $\Delta E$ and the corresponding measured values of desorption energy are shown in Table 1 and Fig 5. The calculated values of $\Delta E$ for 1U-PIM-1+DNT, 2U-PIM+1, 3U-PIM-1+DNT and 10U-PIM-1+DNT are 30.9, 42.6, 68.2 and 75.8 kJ/mol, respectively. The reference point ($\Delta E = 0$) represents a non-interacted case. For comparison we calculated also $\Delta E$ for three configurations of PIM-1 (1U-PIM-1, 2U-PIM-1 and 3U-PIM-1) attached to TNT and BN, respectively. The data in Table 1 shows that the $\Delta E$ results for TNT (31.5, 41.7 and 67.9 kJ/mol) are similar to the corresponding DNT results. In contrast, however, the calculated values of $\Delta E$ for BN are much smaller (7.9, 20.9 and 32.7 kJ/mol) than those of DNT and TNT. The stabilised configuration of the complex indicates that the DNT molecule causes a bending of the rigid planar PIM-1 chain (3U-PIM-1), arising from the Coulomb interaction between the polymer and DNT (see SI 8). We note that the $\Delta E$ value may vary with the relative orientation of 1U-PIM-1 and the interacting molecules. Therefore, we also considered a second possible configuration of 1U-PIM-1 + DNT. The calculated $\Delta E$ for this configuration is 15.9 kJ/mol, compared to 30.9 kJ/mol for the head-to-head configuration. These results indicate that the head-to-head interaction between DNT and 1U-PIM-1 is more stable than the coplanar configuration (SI 9).

In accordance with the discussion above it is clear (see Table 1) that the dominant values of $\Delta E$ are obtained from the configuration of 10U-PIM-1+DNT complex (75.8 kJ/mol). This value is more than twice that of BN complex (32.8 kJ/mol). A detailed analysis of the $\Delta E$ for the 3X-PIM-1+DNT complex reveals that the energy difference between the 2U-PIM-1+DNT and 3U-PIM-1+DNT complexes is 25.6 kJ/mol, whereas the difference between the 3U-PIM-1+DNT and 10U-PIM-1+DNT complexes is 7.6 kJ/mol. These findings indicate that the $\Delta E$ value for the 3U-PIM-1+DNT complex converges more rapidly toward that of the 10U-PIM-1+DNT complex than it does from 2U-PIM-1+DNT to 3U-PIM-1+DNT. The result shows also that the 3U-PIM-1+DNT already brings us close to the experimental value for the binding energy and extending to 10U-PIM-1+DNT makes a further small correction that is much less than the change from 2U-PIM-1+DNT to 3U-PIM-1+DNT. A key finding of this work is that while accurate modelling of binding sites and interaction energies ideally requires a representation that captures the polymeric nature of PIM-1, relatively small molecular fragments can still reproduce the experimentally observed trends in electronic properties with high fidelity. This case study highlights the utility of molecular modelling as a practical tool to strike a balance between computational cost and predictive power, offering valuable guidance for interpreting and designing experiments.

**Table 1** Calculated binding energies ($\Delta E$) for various configurations of XU-PIM-1 interacting with DNT, TNT, and BE. The $\Delta E$ values were computed using B3LYP-gCP-D3/6-31G(d) methods. For comparison, the experimentally determined binding energy of the PIM-1+DNT complex is also included.

| Configurations | ΔE (kJ/mol) B3-LYP-gCP-D3(0)/6-31g(d) | | |
|---|---|---|---|
| | DNT | TNT | BN |
| **Cal. (1U-PIM-1)** | 30.9 | 31.5 | 7.9 |
| **Cal. (2U-PIM-1)** | 42.6 | 41.7 | 20.9 |
| **Cal. (3U-PIM-1)** | 68.2 | 67.9 | 32.7 |
| **Cal. (10U-PIM-1)** | 75.8 | – | 32.8 |
| **Exp. (PIM-1+DNT)** | 73.0 | – | – |

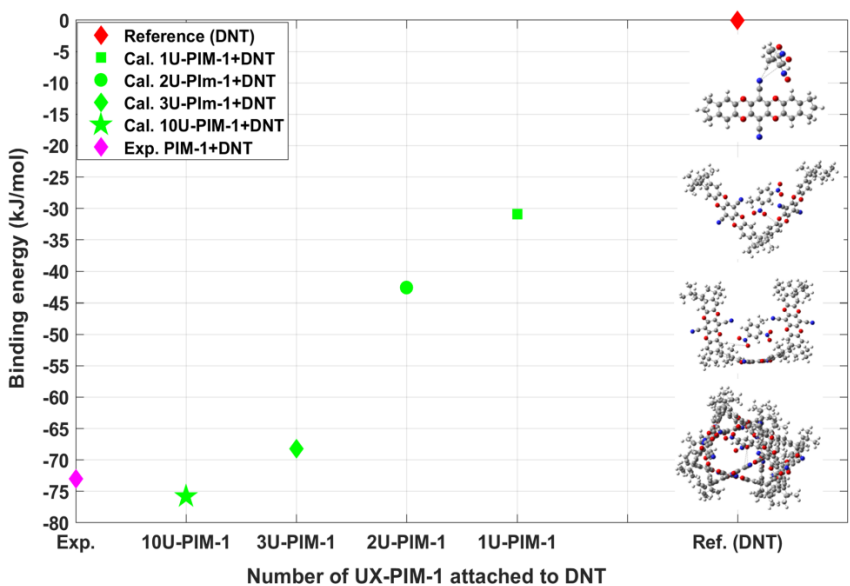

**Figure 5** – Calculated *ΔE* values for the interaction between DNT and four different configurations of PIM-1: 1U-PIM-1+DNT, 2U-PIM-1+DNT, 3U-PIM-1+DNT, and 10U-PIM-1+DNT. The calculation is based on B3LYP-gCP-D3/6-31G(d) functional. Lower *ΔE* values indicate weaker binding interactions. Among the configurations, 10U-PIM-1+DNT exhibits the strongest interaction, with a *ΔE* of 75.8 kJ/mol. Experimental binding energy, estimated from absorbance changes at 241 nm during thermal desorption of DNT, is 73 kJ/mol (indicated in pink), showing an excellent agreement with the predicted value.

**Conclusions**

A motivation for studying explosive analytes at the molecular level is that it renders the possibility to identify the quenching process for detection of explosives substances at ultra-low concentrations. In such applications, changes in the intensity of emission can be used for real-time or off-line sensing. To summarize the finding of the present study, we find that the addition of nitroaromatic analyte leads to a red-shift of the PIM-1 absorption band around 420 nm, and that addition of DNT and TNT causes a greater shift than PIM-1 for PIM-1+BN. The calculated spectra and changes in transition oscillator strength are in good agreement with the corresponding experimental results. The quenching of the emission can be explained by an excited-state electron transfer from PIM-1 to DNT. The calculated frontier molecular orbitals in the composite 1U-PIM+DNT or 1U-PIM+TNT structures show the highest occupied molecular orbital to be located on a planar section of the PIM-1 backbone in each case, while the lowest unoccupied molecular orbital lies on the nitroaromatic analyte. Exposure to DNT or TNT also leads to a strong quenching of the fluorescence, but without significant change in spectrum. These changes relate to a strong non-covalent binding interaction between PIM-1 and DNT or TNT; we calculate the binding energy, using B3LYP-gCP-D3/6-31g(d) functional, to be 75.8 kJ/mol for the interaction of DNT with a 10-unit oligomer of PIM-1. By measuring the changes in absorbance at 230 nm during thermal desorption of DNT, we estimate a binding energy of 73 kJ/mol, in excellent agreement with the calculation.

Molecular interactions typically require a quantum-based approach, which places high demands on both the scalability and accuracy of the applied methodology. In this context, a key finding of this work is that while accurate calculations of molecular binding sites (and their corresponding strengths) generally require models that capture the full polymeric nature of PIM-1, we demonstrate that a smaller molecular model - comprising just a single repeat unit fragment of PIM-1 - can effectively replicate the experimentally observed changes in electronic properties. This case study highlights the potential of molecular modeling to predict experimental outcomes with both flexibility and precision.

These findings indicate also that PIM-1 holds significant promise as a sensor material for explosives. The observed molecular interactions indicate that the analyte molecules dock in a trimer pocket of the highly contorted backbone of the polymer. This could provide valuable insight for developing more sensitive and selective detection systems based on microporous polymers. Continued investigation into these interactions may ultimately enable the design of more advanced sensors capable of distinguishing between various explosive compounds, thereby enhancing both detection accuracy and sensitivity.

## Acknowledgements

S.M. acknowledges financial support from the Swedish Armed Forces (Grant No. FM2022-1661:18) and E.B.O. acknowledges funding from the Commonwealth Scholarship Commission and the Foreign, Commonwealth and Development Office in the UK. This work was supported by a grant from the Engineering and Physical Sciences Research Council (EPSRC) for the project '*Novel Polymers of Intrinsic Microporoisity for use as photonic materials*' (EP/V027840/1, EP/V027735/1, EP/V027425/1). We thank the National Academic Infrastructure for Supercomputing in Sweden (NAISS) at the National Supercomputer Centre of Linköping University partially funded by the Swedish Research Council through grant agreements no. 2022-06725.

# Supporting Information

**Interaction of Polymer of Intrinsic Microporosity PIM-1 with explosive analytes at the molecular level: Combined experiment and computational modelling**

Salam Mohammed[1,2], Edward B. Ogugu[2], Ramakant Sharma[2], Dominic Taylor[3], Graeme Cooke[4], Neil McKeown[3], Glib Baryshnikov[5], Hans Ågren[6,7], Ifor D.W. Samuel[2], Graham A. Turnbull[2]

1. Swedish EOD and Demining Centre-SWEDEC, Swedish Armed Forces, SE-575 28 Eksjö, Sweden.
2. Organic Semiconductor Centre, School of Physics and Astronomy, University of St Andrews, St Andrews, KY16 9SS, UK.
3. School of Chemistry, University of Edinburgh, Edinburgh EH9 3FJ, UK.
4. School of Chemistry, University of Glasgow, Glasgow G12 8QQ, UK.
5. Department of Science and Technology (ITN), Linköping University, SE-581 8 Linköping, Sweden.
6. Department of Physics and Astronomy, X-ray Photon Science, Uppsala University Box 516, 751 20 Uppsala, Sweden.
7. Faculty of Chemistry, Wroclaw University of Science and Technology, Wyspianskiego 27, PL-50370 Wroclaw, Poland

*Author e-mail address: gat@st-andrews.ac.uk and salam.mohammed@mil.se*

## SI. 1

**Synthesis of PIM-1**

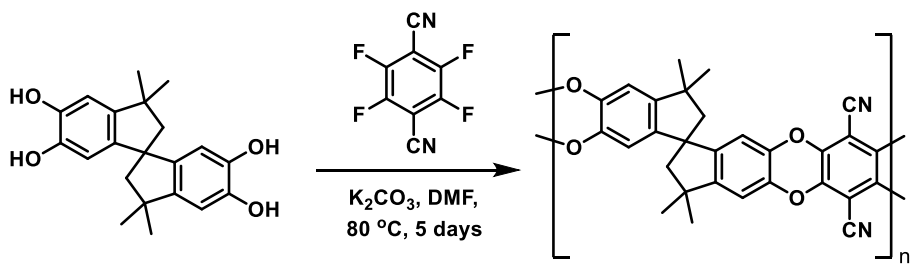

**SI 1** – 5,5',6,6'-Tetrahydroxy-3,3,3',3'-tetramethyl-1,1'-spirobisindane (13.617 g, 40 mmol), tetrafluoroterephthalonitrile (8.004 g, 40 mmol) and potassium carbonate (16.59 g, 120 mmol) were suspended in DMF (150 mL) and heated to 80 °C for 5 days under a nitrogen atmosphere. After cooling, the reaction mixture was poured onto methanol (300 mL) to yield a yellow precipitate that was filtered, washed with water (1 L) and methanol (200 mL). The crude polymer was purified by two reprecipitations from chloroform (300 mL) solution into methanol (1 L). The crude product was then dried in a conventional oven for 3 days at 100 °C then a vacuum oven at 100 °C for 6 hours to yield a bright yellow powder (17.02 g, 92%). $^1$H NMR (500 MHz, CDCl$_3$, 25.0 °C) $\delta_H$ 6.81 (bs, 2 *H*), 6.41 (bs, 2 *H*), 2.34 (bs, 2 *H*), 2.16 (bs, 2 *H*), 1.56 (s, 6 *H*), 1.37 (s, 6 *H*), 1.31 (s, 6 *H*). Data in agreement with reported spectra. Gel permeation chromatography, eluent = chloroform, calibrated against polystyrene standards: $M_W$ = 104 870 g mol$^{-1}$, $M_n$ = 17 711 g mol$^{-1}$, PDI = 5.92

**SI 2.**

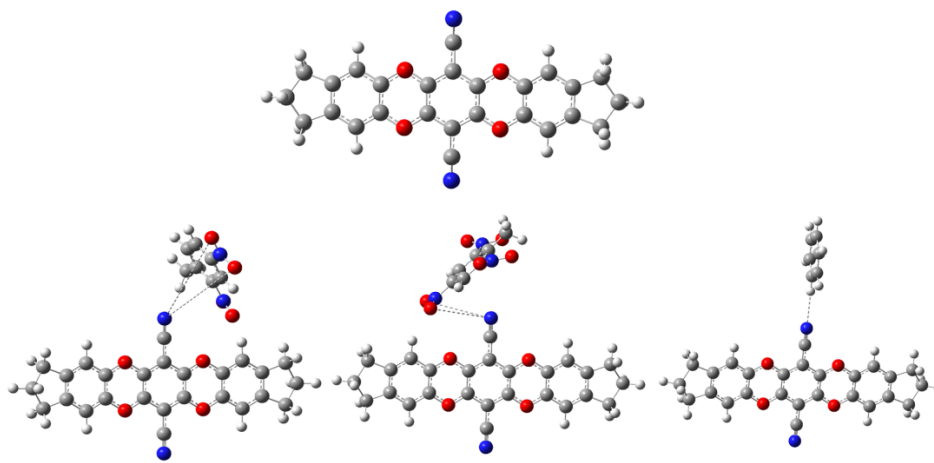

**SI 2** – Optimized molecular structure (head-to-head) interaction between the CN group of 1U-1U-PIM-1 and analytes of DNT, TNT and BN, respectively.

**SI 3.**

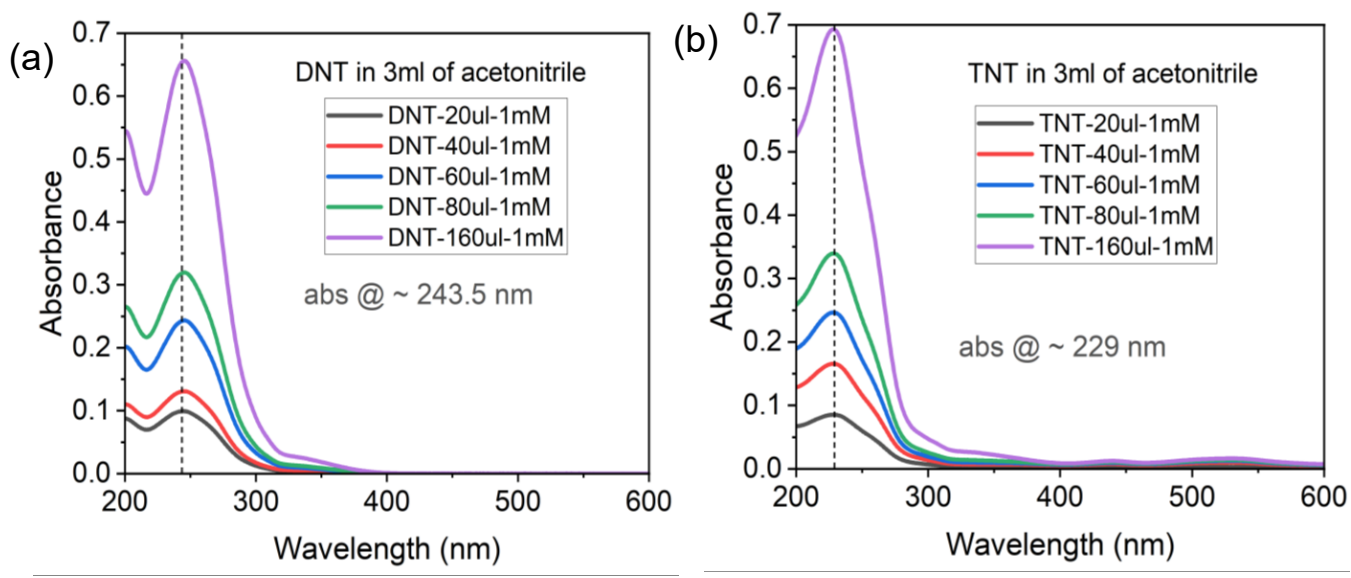

**SI 3** – Optical absorption of 2,4-DNT and 2,4,6-TNT in acetonitrile solutions: (a) absorbance increase with increase in DNT concentration. (b) increase in absorbance with increasing TNT concentration.

A stock solution of DNT was prepared at a concentration of 1 mM in acetonitrile. Then, 20 µL of the 1 mM DNT solution was added to 3 mL of acetonitrile in a cuvette before absorption measurement. The procedure of adding DNT solution was repeated up to 5 aliquots, as shown in

SI 3(a), yielding DNT concentrations of 3.64 µg, 7.28 µg, 10.92 µg, 14.56 µg, and 29.12 µg, respectively. A similar experiment was conducted for the TNT solution, as shown in SI 3(b), yielding concentrations of 4.54 µg, 9.09 µg, 13.63 µg, 18.17 µg, and 36.34 µg, respectively.

**SI 4**

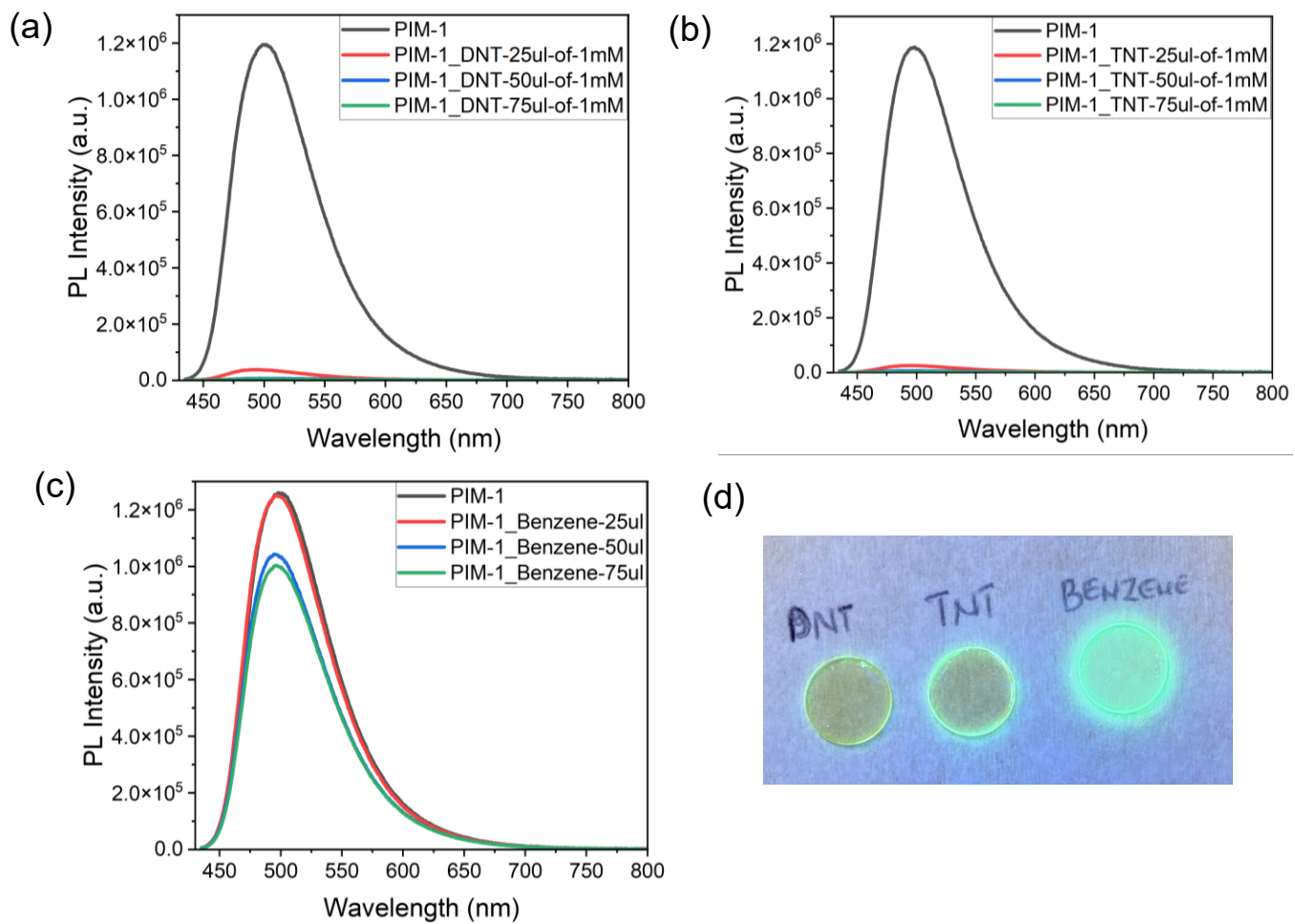

**SI 4** – Photoluminescnce (PL) of PIM-1 and subsequent PL quenching due analytes: (a) PL quenching of PIM due to DNT at various concentration, (b) PL quenching with increasing concentration of TNT, (c) Slight drop in the PL of PIM-1 due to benzene solvent, and (d) a photo showing PIM-1 films doped with 13.66 µg of DNT, 17.03 µg of TNT, and 75 µL of benzene.

The PIM-1 films were doped with 2,4,6-trinitrotoluene (TNT) and 2,4-dinitrotoluene (DNT) by drop-casting 25 µL of the respective solutions of 1 mM in acetonitrile onto the PIM-1 films, and solutions were allowed to evaporate, leaving behind molecules of analytes adsorbed in the films before PL measurements. The response of PIM-1 to non-explosive molecules was also explored by drop casting 25 µL of benzene. The procedure was repeated up to three aliquots to examine the concentration of analytes, which would result in complete PL quenching of the PIM-1. The three aliquots employed resulted in concentrations of 4.55 µg, 9.11 µg, and 13.66 µg, respectively, of DNT in PIM-1, and 5.68 µg, 11.36 µg, and 17.03 µg, respectively, of TNT in PIM-1.

**SI 5.**

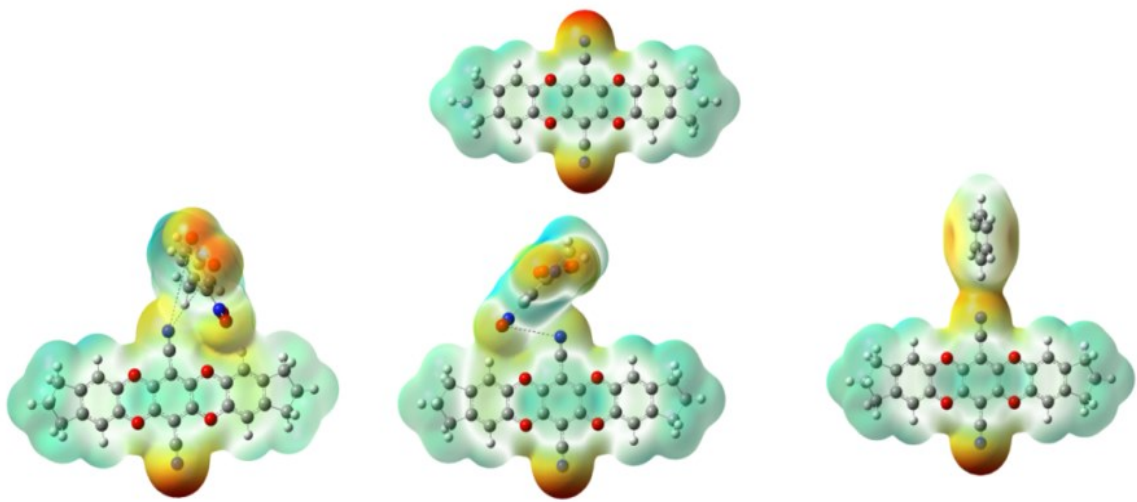

**SI 5** – Molecular electrostatic potentials (MEP) maps of 1U-PIM-1 (upper) and Interactions between the CN group of 1U-PIM-1 with the analyte of DNT, TNT and BN, respectively (lower). Electron rich (negative charge) and the electron deficient (positive charge) is represented by red and blue color, respectively.

**SI. 6**

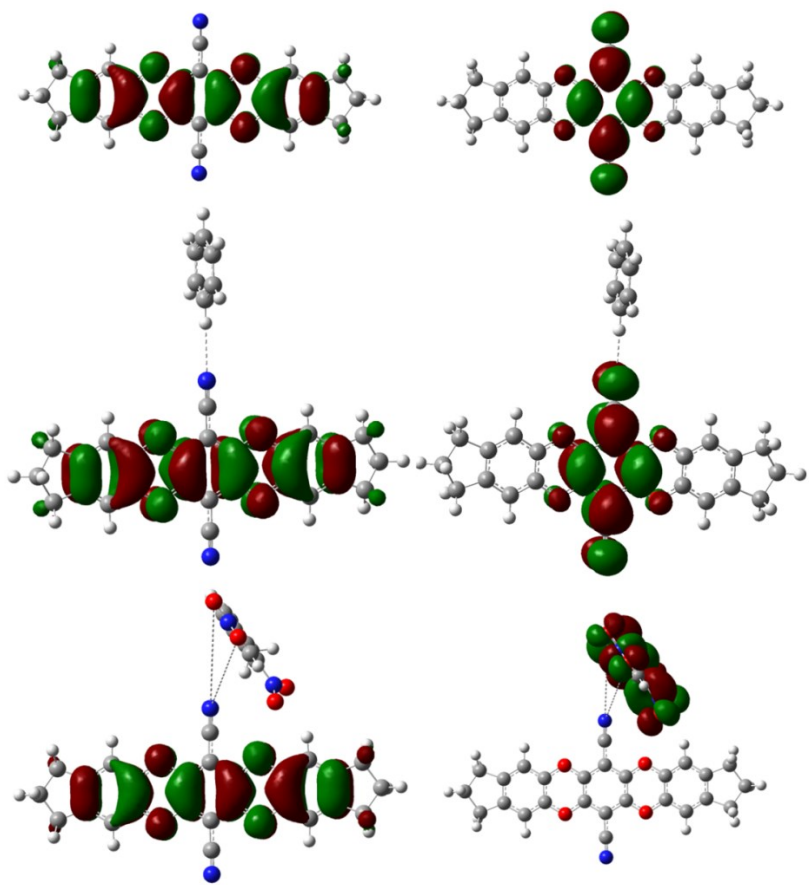

**SI 6** – Calculated frontier molecular orbitals (HOMO and LUMO) in the composite 1U-PIM-1 (Upper), 1U-PIM+BN (Meddle) and 1U-PIM+DNT (lower). The structures show the highest occupied molecular orbital HOMO to be located on a planar section of the 1U-PIM-1 backbone in each case, while the lowest unoccupied molecular orbital LUMO lies on the nitroaromatic analyte of DNT as the only case.

**SI 7.**

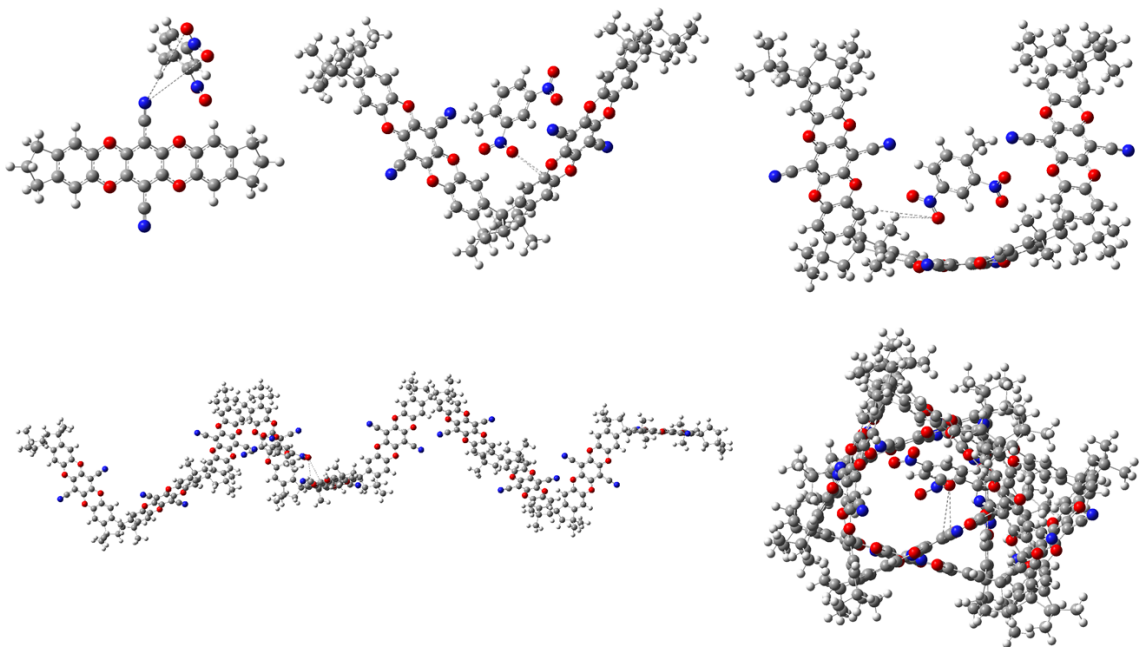

**SI 7** – Four different optimized configurations were used for the calculations of the binding energy $\varDelta E$ between PIM-1 and DNT. 1U-PIM-1+DNT (upper-left), 2U-PIM-1+DNT (upper-middle), 3U-PIM-1+DNT (upper-right) and 10U-PIM-1+DNT (lower-left: front view and lower right: the side view).

**SI 8.**

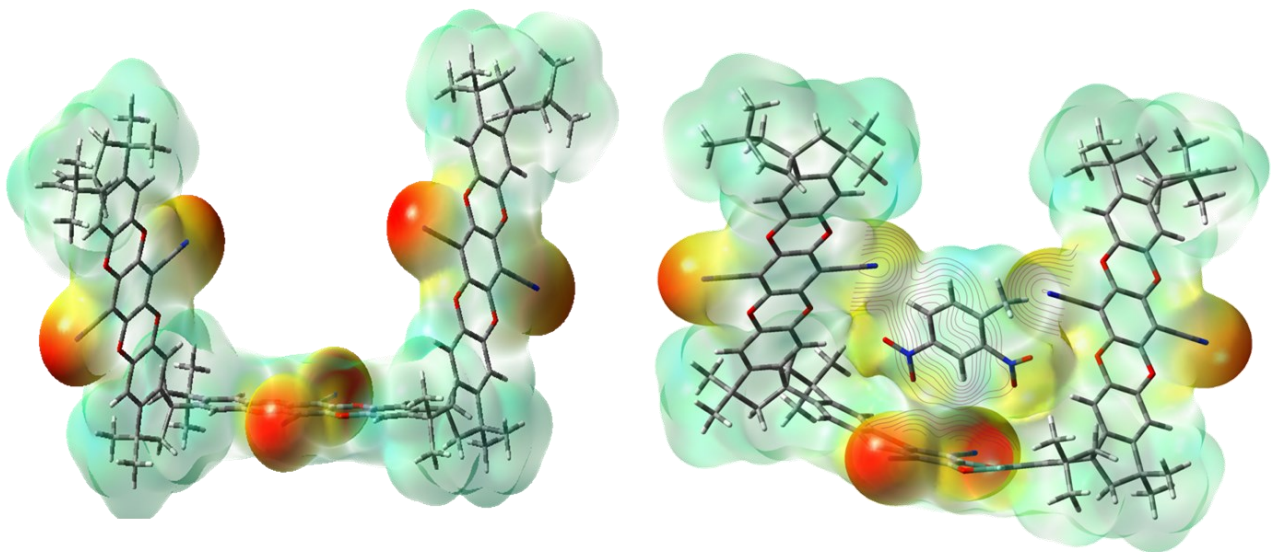

**SI 8** – Molecular electrostatic potentials (MEP) maps show the DNT molecule causes a bending of the rigid planar PIM-1 chain 3U-PIM-1, arising from the coulomb interaction between the polymer and DNT (compare left: 3UPIM-1 to right:3U-PIM+DNT). Electron rich (negative charge) and the electron deficient (positive charge) is represented by red and blue color, respectively.

**SI 9.**

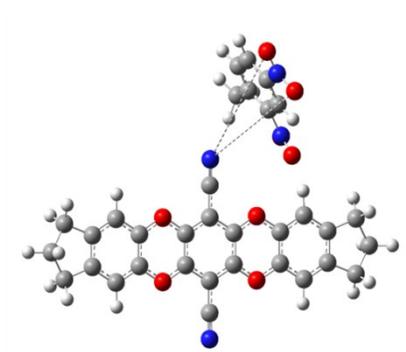

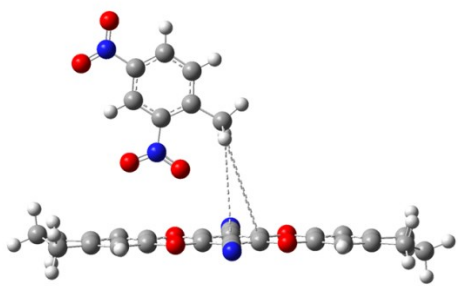

**SI 9** – Calculated *ΔE* for head-to-head (upper) and coplanar (lower) of 1U-PIM-1+DNT is 30.9 and 15.9 kJ/mol, respectively.

## SI. 10

**Calculated the *ΔE* (kJ/mol) for 3U-PIM+DNT using B3LYP-D3-gCP/6-31G(d)/ 6-31g(d)**

**DNT: (a.u.)**

SCF energy         -680.561825657

gCP correction      0.0456961305

D3 correction      -0.01627335

gCP-D3 correction   0.0294227805

**SCF-gCP-D3 energy -680.5324028765**

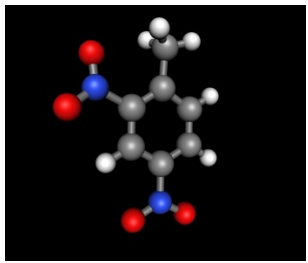

**3U-PIM (a.u.)**

SCF energy         -5089.88617953

gCP correction      0.4602602174

D3 correction      -0.25417230

gCP-D3 correction   0.2060879174

**SCF-gCP-D3 energy -5089.6800916126**

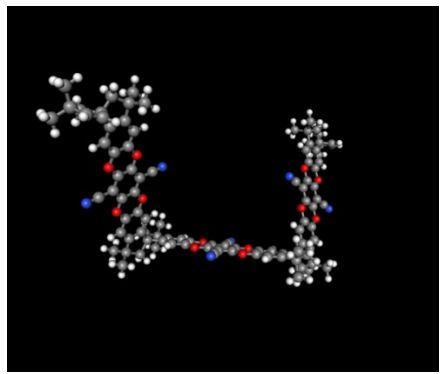

**3U-PIM+DNT (a.u)**

SCF energy         -5770.46445349

gCP correction      0.5169216454

D3 correction      -0.29092635

gCP-D3 correction   0.2259952954

**SCF-gCP-D3 energy  -5770.2384581946**

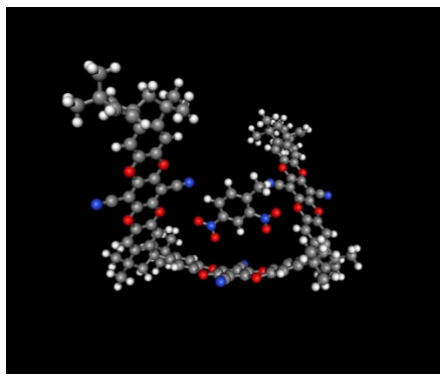

**Cal. (*ΔE*) $_{3U\text{-}PIM+DNT}$ = - 68.2 kJ/mol,**

**Exp. (*ΔE*) $_{PIM\text{-}1+DNT}$ = 73 kJ/mol**

**Calculated the *ΔE* (kJ/mol) for 3U-PIM+TNT using B3LYP-D3-gCP/6-31G(d)/ 6-31g(d)**

<u>TNT (a.u.):</u>

SCF energy           -885.045483764

gCP correction       0.0545468025

D3 correction        -0.02132730

gCP-D3 correction    0.0332195025

**SCF-gCP-D3 energy -885.0122642615**

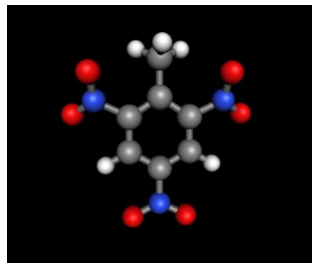

<u>3U-PIM (a.u.)</u>

SCF energy           -5089.88617953

gCP correction       0.4602602174

D3 correction        -0.25417230

gCP-D3 correction    0.2060879174

**SCF-gCP-D3 energy -5089.6800916126**

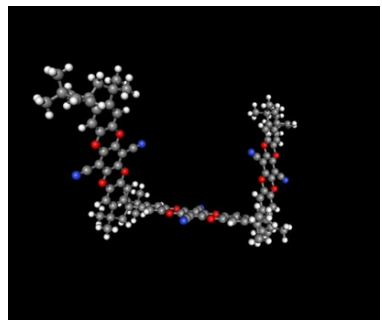

<u>3U-PIM-1+TNT(a.u.)</u>

SCF energy           -5974.94746478

gCP correction       0.5261230893

D3 correction        -0.29686098

gCP-D3 correction    0.2292621093

**SCF-gCP-D3 energy -5974.7182026707**

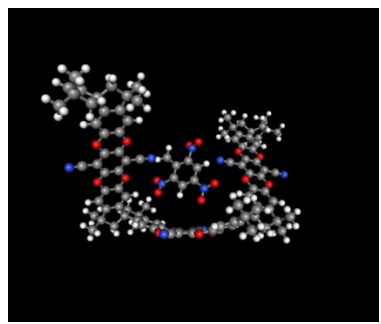

**Cal. (ΔE) 3U-PIM+TNT = - 67.86 kJ/mol**

**Cal. (*ΔE*) 3U-PIM+DNT = - 68.2 kJ/mol,**

**Exp. (*ΔE*) PIM-1+DNT = 73 kJ/mol**

**Calculated the *ΔE* (kJ/mol) for 3U-PIM+BN using B3LYP-D3-gCP/6-31G(d)/ 6-31g(d)**

**BN:**

SCF energy    -232.248649431

gCP correction 0.0220835213

D3 correction -0.00509906

gCP-D3 correction 0.0169844613

**SCF-gCP-D3 energy -232.2316649697**

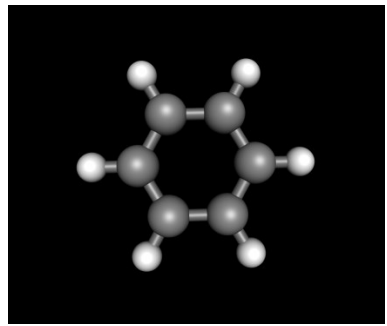

**3U-PIM**

SCF energy   -5089.88617953

gCP correction 0.4602602174

D3 correction -0.25417230

gCP-D3 correction    0.2060879174

**SCF-gCP-D3 energy -5089.6800916126**

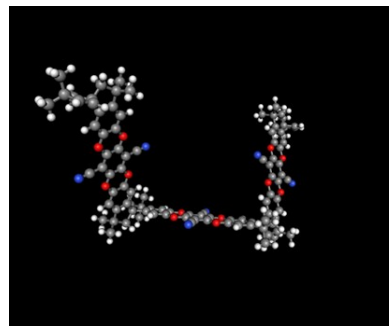

**3U-PIM+BN**

SCF energy   -5322.13887480

gCP correction 0.4848322965

D3 correction -0.27018913

gCP-D3 correction 0.2146431665

**SCF-gCP-D3 energy -5321.9242316335**

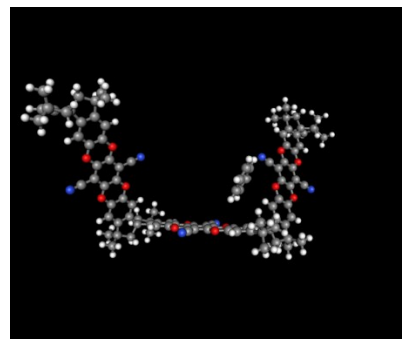

**Cal. (*ΔE*) 3U-PIM+BN = -32.75 kJ/mol**

**Cal. (ΔE) 3U-PIM+TNT = - 67.86 kJ/mol**

**Cal. (*ΔE*) 3U-PIM+DNT = - 68.2 kJ/mol,**

**Exp. (*ΔE*) PIM-1+DNT = 73 kJ/mol**

## Calculated the BE of $\Delta E$ (kJ/mol) for 10U-PIM+DNT using B3LYP-D3-gCP/6-31G(d)/ 6-31g(d)

**DNT (a.u.):**

SCF energy         -680.561825657

gCP correction       0.0456961305

D3 correction       -0.01627335

gCP-D3 correction    0.0294227805

**SCF-gCP-D3 energy -680.5324028765**

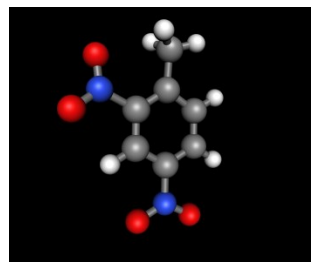

**10U-PIM(a.u.)**

SCF energy         -15782.347172
gCP correction 1.3589258086
D3 correction  -0.7773596
gCP-D3 correction     0.5815662086
**SCF-gCP-D3 energy -15781.7656057914**

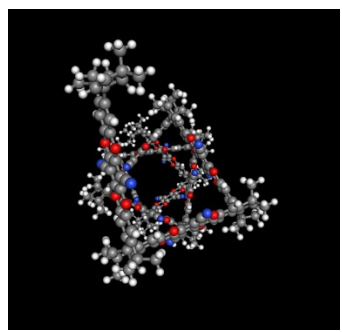

**10U-PIM+DNT(a.u.)**

SCF energy       -16462.922157983
gCP correction      1.4193129541
D3 correction       - 0.8240356170
gCP-D3 correction   0.5952773371
**SCF-gCP-D3 energy   -16462.3268806459**

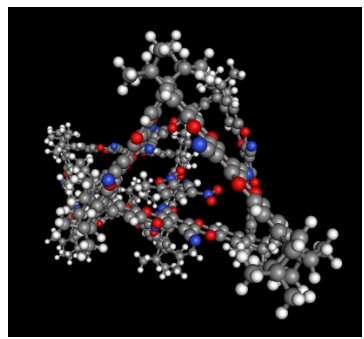

**Cal. ($\Delta E$) $_{10U-PIM-1+DNT}$ = - 75.8 kJ/mol,**

**Exp. ($\Delta E$) $_{PIM-1+DNT}$ = 73 kJ/mol**

# Calculated the BE of $\Delta E$ (kJ/mol) for 10U-PIM+BN using B3LYP-D3-gCP/6-31G(d)/ 6-31g(d)

**BN:**

SCF energy   -232.248649431

gCP correction 0.0220835213

D3 correction -0.00509906

gCP-D3 correction 0.0169844613

**SCF-gCP-D3 energy -232.2316649697**

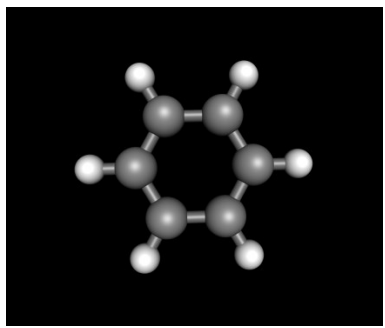

**10U-PIM(a.u.)**

SCF energy        -15782.347172
gCP correction      1.3589258086
D3 correction       -0.7773596
gCP-D3 correction    0.5815662086
**SCF-gCP-D3 energy -15781.7656057914**

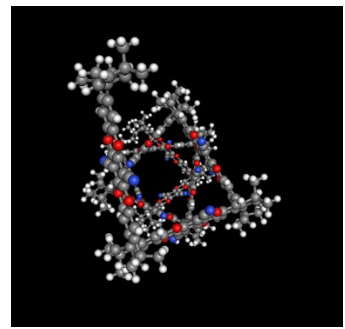

**10U-PIM+BN (a.u.)**

SCF energy        -16015.3910506  -16,014.6048629003

gCP correction      1.3812964564

D3 correction      -0.7861876997

gCP-D3 correction    1.3812964564

**SCF-gCP-D3 energy -16014.0097541436**

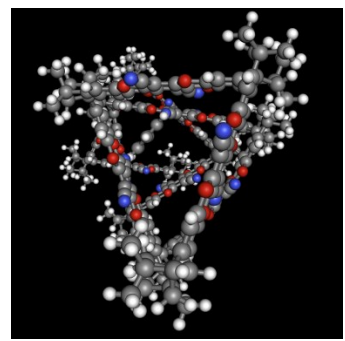

**Cal. ($\Delta E$) $_{10U\text{-}PIM\text{-}1+BN}$ = -32.8 kJ/mol**

**Cal. ($\Delta E$) $_{10U\text{-}PIM\text{-}1+DNT}$ = - 75.8 kJ/mol,**

**Exp. ($\Delta E$) $_{PIM\text{-}1+DNT}$ = 73 kJ/mol**

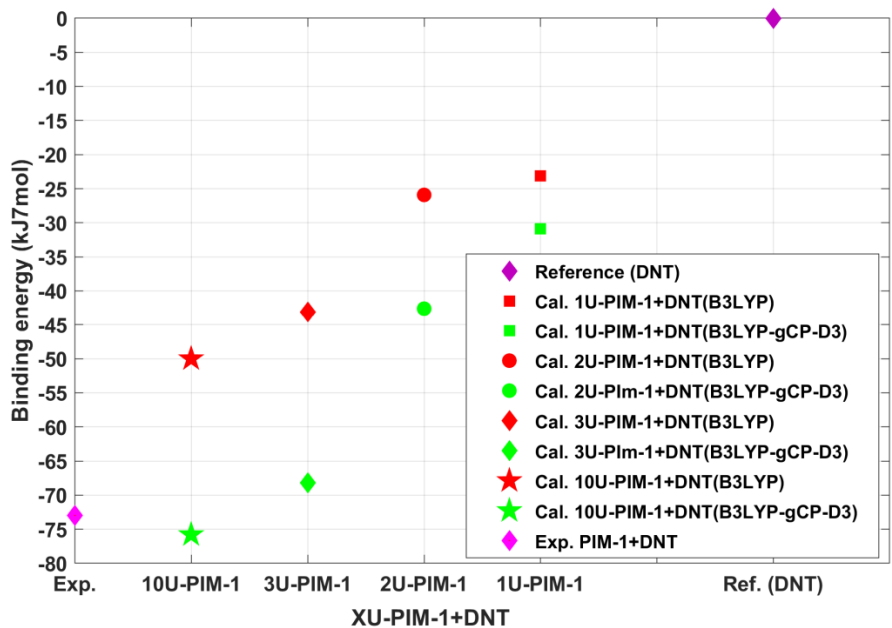

**SI 10** – Calculated $\Delta E$ values for the interaction between DNT and four different configurations of PIM-1: 1U-PIM-1+DNT, 2U-PIM-1+DNT, 3U-PIM-1+DNT, and 10U-PIM-1+DNT. Lower $\Delta E$ values indicate weaker binding interactions. Results from standard B3LYP calculations are shown in red, while values obtained using B3LYP with gCP-D3 corrections are shown in green. Among the configurations, 10U-PIM-1+DNT exhibits the strongest interaction, with a $\Delta E$ of 75.8 kJ/mol. Experimental binding energy, estimated from absorbance changes at 241 nm during thermal desorption of DNT, is 73 kJ/mol (indicated in pink), showing good agreement with the corrected theoretical value.

**S10** Calculated binding energies ($\Delta E$) for various configurations of XU-PIM-1 interacting with DNT, TNT, and BE. The $\Delta E$ values were computed using both B3LYP/6-31G(d) and B3LYP-gCP-D3/6-31G(d) methods. For comparison, the experimentally determined binding energy of the PIM-1+DNT complex is also included.

| Configurations | $\Delta E$ (kJ/mol) B-3LYP/B3-LYP-gCP-D3(0)/6-31g(d) | | |
|---|---|---|---|
| | DNT | TNT | BN |
| Cal. (1U-PIM-1) | 23.1/ 30.9 | 23.1/ 31.5 | 5.8/ 7.9 |
| Cal. (2U-PIM-1) | 25.9/ 42.6 | 25.3/ 41.7 | 6.1/ 20.9 |
| Cal. (3U-PIM-1) | 43.1/ 68.2 | 41.5/ 67.9 | 10.6/ 32.7 |
| Cal. (10U-PIM-1) | 50.0/75.8 | 45.2/ | 25.5/32.8 |
| Exp. (PIM-1+DNT) | 73.0 | — | — |